\documentclass[conference]{IEEEtran}
\usepackage{xcolor}
\usepackage{cite}
\usepackage{amsmath,amssymb,amsfonts}
\usepackage{graphicx}
\usepackage{textcomp}
\usepackage{colortbl}
\usepackage{array}
\usepackage{multirow}
\usepackage{booktabs}
\usepackage{xurl}
\usepackage{tikz}
\usetikzlibrary{arrows.meta,positioning,calc}
\usepackage[hidelinks]{hyperref}
\hypersetup{pdftitle={Local Is Not a Sufficient Privacy Boundary: Governing OS-Integrated On-Device AI},pdfauthor={Jonghyun Chung and Sanket Badhe}}

\newcolumntype{P}[1]{>{\raggedright\arraybackslash}p{#1}}
\newcolumntype{D}[1]{>{\columncolor{black!5}\raggedright\arraybackslash}p{#1}}

\tikzset{
  diagrambox/.style={
    draw,
    rounded corners=5pt,
    semithick,
    align=center,
    inner xsep=7pt,
    inner ysep=6pt,
    fill=white
  },
  workflowbox/.style={
    diagrambox,
    text width=0.79\columnwidth,
    minimum height=0.95cm,
    fill=black!3
  },
  diagramarrow/.style={
    -{Latex[length=2.2mm]},
    semithick
  }
}

\title{Local Is Not a Sufficient Privacy Boundary: Governing OS-Integrated On-Device AI}
\author{\IEEEauthorblockN{Jonghyun Chung and Sanket Badhe}
\IEEEauthorblockA{Google LLC\\
johnchung@google.com; sanketbadhe@google.com}}

\begin{document}
\maketitle

\begin{abstract}
On-device inference limits remote exposure but does not establish privacy when an operating-system assistant assembles cross-application data, retains derived state, or delegates actions. We present a feature-boundary audit method combining six privacy-failure surfaces, a five-dimension rubric with separate evidence grades, and records for adjacent data and authority paths. We apply it to Apple Foundation Models, Apple Private Cloud Compute, Android AICore/Gemini Nano, and Microsoft Recall using primary public documentation. A formative comparison of two 20-item assessment records identifies ambiguities in notice, retention, path ownership, and accountability. Across the four boundaries, D1 and D2 reach level~1 while D4 remains undetermined, reflecting coarse documented boundaries but unresolved workflow mediation. We also specify a minimized event record and release comparison for fleet security triage and scoped forensic review.
\end{abstract}


\section{Introduction}
\label{sec:introduction}

Artificial intelligence (AI) is moving from application features into operating systems (OSs). Apple, Google, and Microsoft document local model runtimes, device APIs, assistant surfaces, semantic indexes, and private-cloud paths near user content \cite{apple_foundation_models, android_gemini_nano, microsoft_recall_manage}. The security and privacy unit of analysis therefore shifts from a model or narrow API to a mediation layer that selects context, retains derived state, authorizes actions, routes data, and changes after updates.

Local inference reduces routine remote transfer but answers only where computation occurs. It does not determine who assembles context, what derived state persists, what actions disclose, how fallback is reported, or how updates change authority. A local meeting assistant can still combine calendar metadata, email, drafts, and inferred facts. The privacy risk is mediated information flow, not model inference alone.

Across managed devices, the provenance, retention, and recipients of screenshots, messages, indexes, and event streams also determine whether release changes are comparable. Section~\ref{sec:fleet} specifies a feature-boundary event record for analyst triage and forensic review.

We therefore audit OS-integrated AI at the \emph{feature boundary}: an auditable capability or workflow and its declared sources, retrieval, model calls, memory, tools, routes, logs, and update dependencies. Excluded host or platform paths must be registered as adjacent boundaries, and a component label cannot be generalized to an end-to-end product.

Our contribution is a structured synthesis and audit-record discipline that re-specifies least privilege, contextual integrity, minimization, retention limitation, provenance, and auditability for OS mediation. Versioned records expose adjacent-path dependencies and claim-specific evidence; for managed fleets they also group minimized events by capability, policy, and release. Evidence is classified by what it demonstrates, not whether it is public. The obligations and evidence are primary; a numeric label is a lossy summary.

The paper develops an OS-centered threat model, a six-surface taxonomy, and a four-level rubric, then applies them to three component boundaries---Apple Foundation Models, Apple Private Cloud Compute, and Android AICore/Gemini Nano---and the Microsoft Recall snapshot/index workflow. A formative two-record comparison surfaces ambiguities in ownership, notice, retention, inference restraint, and accountability.

\section{Background and Gap}
\label{sec:background}

Contextual integrity and privacy taxonomies treat privacy as constrained information flow rather than secrecy alone \cite{nissenbaum_context, solove_taxonomy}. Privacy-by-design, LINDDUN, DPIA practice, and privacy-risk frameworks translate that concern into minimization, purpose limitation, retention, transparency, and accountability requirements \cite{cavoukian_pbd, deng_linddun, hoepman_privacy_design_strategies, gdpr_article35, ico_dpia_guidance, nist_privacy_framework}. Classic platform-security principles motivate least privilege, complete mediation, and confused-deputy defenses \cite{saltzer_schroeder, hardy_confused_deputy}. Mobile permission and information-flow research shows why consent and sandboxing are incomplete once data moves across app and OS mediation layers \cite{felt_android_permissions, roesner_user_driven_access, enck_taintdroid}.

Research on voice assistants identifies bystander, invocation, skill-confusion, and user-understanding risks in ambient systems \cite{lau_alexa_listening, malkin_smart_speakers, abdi_smart_speakers, zhang_dangerous_skills}. Work on model and embedding privacy shows that derived representations may disclose sensitive content and should not be treated as harmless metadata \cite{song_embedding_leakage, morris_embeddings_reveal, carlini_extracting, nasr_privacy_dnn}. Agent-security research adds indirect prompt injection, memory, access-control, and tool-use failures \cite{greshake_indirect_prompt_injection, wang_memory, shi_tool_selection, doshi_safe_tool_use, li_access_control}. Governance and audit research adds transparency, contestability, risk management, and post-deployment review \cite{nist_ai_rmf, nist_genai, raji_gap, diakopoulos_accountability}.

Recent agent-privacy work moves beyond final-output leakage. A data-centric survey organizes risk across queries, intermediate results, memory, tools, and messages; AgentLeak instruments internal coordination channels; OCELOT treats inference leakage across an agent trajectory; and Aura proposes a mediated mobile-agent OS architecture \cite{lahjouji_agents_too_much, elyagoubi_agentleak, xie_ocelot, zou_blind_gods}. PrivacyLens evaluates contextual privacy norms along agent trajectories, AgentDojo evaluates prompt injection in tool-using agents, and GAAP enforces information-flow permissions across agent steps and persistent stores \cite{shao_privacylens, debenedetti_agentdojo, stanley_gaap}. Other assistant evaluations operationalize contextual integrity for form filling, test contextual flows over synthetic communications, or measure task-irrelevant data use by web agents \cite{ghalebikesabi_operationalizing_ci, cheng_cibench, zharmagambetov_agentdam}.

AgentSCOPE is the closest point of comparison. Its Privacy Flow Graph uses contextual-integrity parameters to evaluate intermediate information flows in multi-tool agent workflows \cite{ngong_agentscope}. Where AgentSCOPE evaluates information-flow edges within an agent trajectory, our unit is a versioned OS capability boundary that also covers persistent derived state, co-user and administrator access, private-cloud and telemetry paths, composition across adjacent boundaries, and replay after platform change. A permission may allow calendar access without constraining whether an assistant combines that event with email, stores an inferred fact, or later places it in a message. Our synthesis connects those OS-level decisions to evidence requirements and scoring rules at the feature boundary.

The evidence discipline also draws on security assurance: Common Criteria separates a Target of Evaluation from its environment and treats composition explicitly, NIST SP~800-53A scopes assessment methods to selected controls, and service-organization reports retain user and subservice dependencies \cite{common_criteria_2022,nist_sp800_53a,apple_pcc_soc3_report}. Like Goal Structuring Notation assurance cases, our records separate claims, context, evidence, and unresolved dependencies \cite{gsn_standard}. We specialize those ideas to derived memory, delegated authority, adjacent routes, and OS change replay rather than constructing a complete system safety argument.

Architectural risk analysis, asset-centric AI threat modeling, MITRE ATLAS, DASF, and the NIST AI RMF already provide system inventories, adversary techniques, and risk-management processes \cite{mcgraw_ara,vicarte_asset_threat,mitre_atlas,databricks_dasf,nist_ai_rmf}. Our method attaches a capability record to those analyses: it names raw and derived state, local access classes, tool and remote recipients, evidence for each control claim, and the changes that trigger re-evaluation. It does not replace threat selection or estimate attack likelihood.

The resulting contribution combines a versioned OS capability boundary, adjacent-path closure, claim-specific evidence grades, and replay after platform change. Unlike task-trajectory analyses, it also covers ambient indexes, co-users and administrators, persistent derived state, fallback, telemetry, and OS updates. Isolation and attestation evidence still does not establish task-appropriate data use.

\section{Method and Scope}

This paper combines structured conceptual synthesis, bounded analysis of public platform documentation, and a formative comparison of two assessment records. The unit of analysis is a document-backed \emph{feature boundary}: a publicly described platform capability or user-visible workflow, its declared technical scope, and the claims about its context sources, retained state, tool or fallback paths, user controls, administrative controls, and update governance. The platform application is documentary; the implementation probes are specifications for future execution rather than measurements reported here. Section~\ref{sec:formative} reports the assessment comparison separately from the platform corpus.

We used a conceptual corpus and a platform corpus. The former covers privacy engineering, AI threat modeling, OS permissions, assistant privacy, and audit; it grounds the method rather than vendor findings. The platform corpus consists of primary materials describing Foundation Models, Apple Intelligence and Private Cloud Compute (PCC), AICore/Gemini Nano and Android Private Compute Core, and Recall \cite{apple_foundation_models, apple_wwdc25_foundation_models, apple_wwdc26_foundation_models, apple_intelligence_privacy, apple_foundation_models_policy, apple_pcc_security, apple_pcc_security_guide, android_gemini_nano, android_aicore_privacy, android_pcc_architecture, google_android_aicore_help, microsoft_recall_manage, microsoft_recall_click_to_do, microsoft_recall_privacy, microsoft_recall_architecture, microsoft_recall_snapshots_disk}. The Foundation Models and Recall records retain their July 24, 2026 cutoff. The AICore boundary includes the Gemini Nano documentation revision dated September 8, 2026, and a targeted PCC review runs through September 14, 2026. These cutoffs define separate historical profiles, not a contemporaneous vendor ranking. The supplementary 20-cell evidence ledger gives the four control profiles' cell-level basis, while Section~\ref{sec:pcc} separately examines PCC agent paths and provisioning assurance. The supplement records these cutoffs, source locators, and access dates.

The cases purposively span two local-model APIs, one private-cloud path, and one ambient capture/index workflow. Selection illustrates different boundary types, rather than comparing matched vendor offerings. End-to-end assistants and Google's Private AI Compute are outside these four cases; the latter has a public architecture and a scoped NCC Group review \cite{google_private_ai_compute,ncc_private_ai_compute}. Its omission limits cross-vendor conclusions, not the availability of evidence. Adding it after refinement on the four cases would not constitute a held-out test; a transfer study should freeze the rubric and case protocol before scoring it. Throughout, PCC denotes Apple's Private Cloud Compute; Android Private Compute Core is written in full.

We included public, attributable material that directly described an evaluated feature boundary or audit dimension. We excluded press coverage, forum posts, authentication-gated material, and documents that did not describe the boundary being scored. The final manifest contains 24 included platform entries; selection was purposive rather than exhaustive, so it does not support a systematic-review coverage claim. For each platform source, we recorded its issuer, title, URL, access date, feature boundary, evidence class, extracted claim or control, rubric dimension, and a page, heading, or stable-URL locator. Academic and governance sources inform the framework; primary platform documents and the published independent examination support the platform evidence records.

The synthesis treated context assembly, memory persistence, delegated action, telemetry, fallback routing, and post-deployment change as sensitizing categories rather than fixed codes. We retained a concept when it identified an asset, a trust boundary, a privacy failure, and a possible control point. For example, indirect prompt injection identifies untrusted retrieved content, a failure in instruction authority, disclosure or tool misuse, and controls at provenance labeling and tool mediation. Concepts without an auditable control point remain background rather than rubric dimensions.

We recorded source genre separately from evidentiary content: policy, API or architecture description, inspectable code, execution record, or completed evaluation. Public availability does not determine the evidence grade. We inspected platform documents and the published scope of a PCC provisioning examination. A report schema and a collected execution trace therefore receive different evidence grades; scope statuses are assigned under the boundary discipline below.

\subsection{Boundary Discipline}
Before scoring, the evaluator records the task, version, account/region, policy state, sources, stores, tools, routes, telemetry, and updates. A subclaim is material when reversing its disposition could change an in-scope claim or probe outcome by changing permitted sources or recipients, durable state, user notice or agency, or review/update obligations. This decision is recorded before outcome evidence is examined; uncertain materiality remains UD. N/A requires evidence that a path is absent; an applicable but unresolved path is UD; a reachable path assigned to a named boundary is ADJ. ADJ and N/A ordinarily classify paths, not whole dimensions. Once all material subclaims are evaluated, the workflow label is the lowest supported applicable anchor across them: an absent path drops out, a material UD propagates UD, and a material ADJ blocks composition until its target is evaluated. Mixed internal and adjacent obligations otherwise remain UD. Component scores cannot be generalized to the complete product.

We distinguish five feature types because they require different evidence: local model APIs, ambient semantic indexes, assistant/tool layers, private-cloud routes, and enterprise-managed capture/search. A product can combine them, but a score for one type cannot be inherited by another.

\section{Threat Model and Privacy Failure Surfaces}

We model each capability as a feature-boundary instance: a user task plus the sources, derived state, actions, fallback paths, and policy dependencies needed to complete it. Figure~\ref{fig:systemmodel} shows five control points. C1 governs source eligibility; C2 applies task purpose, provenance, domain labels, redaction, and instruction hierarchy; C3 covers model, retrieval, and memory state; C4 mediates tools, telemetry, and fallback; and C5 covers debugging, updates, feature flags, administrative policy, and release review.

The principals are the user or session owner, local co-users, third-party apps, authors of retrieved content, the platform provider, enterprise or parental administrators, tool recipients, fallback/cloud operators, and auditors. Assets include raw content, metadata, prompts, retrieved snippets, provenance labels, embeddings, summaries, inferred facts, durable memories, tool arguments, drafts, telemetry, fallback requests, logs, policy state, and model artifacts. These states are not equivalent: embeddings, summaries, and inferred facts are derived privacy objects rather than anonymous implementation metadata \cite{song_embedding_leakage, morris_embeddings_reveal}.

The platform provider is a \emph{semi-trusted} governance actor. A fully malicious OS provider can read local files, observe memory, ship a hostile update, and forge user-facing evidence; that case reduces to general OS compromise. Our model instead places provider components in the trusted computing base for mediation and updates while treating their defaults, documentation, policy choices, telemetry, routing, and later changes as claims to audit. The provider is assumed not to deliberately exfiltrate all data, but may still create failures through broad defaults, opaque fallback, negligent retention, compromised services, administrator delegation, or governance drift.

Adversaries may place instructions in documents, web pages, email, messages, calendar fields, screenshots, or retrieved media; build apps that exploit broad assistant access; induce a confused-deputy action; inspect local AI state as a co-user, administrator, repair/forensic actor, or malware process within the evaluated access class; receive overbroad tool outputs; or benefit from fallback and update changes. Prompt injection crosses C1--C4; cross-context leakage crosses C2--C3; memory poisoning and persistence concentrate at C3 and C5; tool exfiltration crosses C4; co-user and administrator access crosses C1, C3, and C5; and fallback, telemetry, and update drift cross C4--C5.

Each audit states its access class: A0, a content author who plants instructions or facts; A1, a host or third-party app with ordinary permissions; A2, a tool recipient or external service; A3, a local co-user; A4, an administrator, repair, or forensic actor; A5, a fallback or telemetry operator observing routed data; or A6, local malware within a declared non-root privilege class. A result for one class does not imply a stronger class. Non-goals are a provider that intentionally forges evidence or exfiltrates all data, kernel/root compromise, hardware side channels, formal verification of closed components, model-quality evaluation, and legal-compliance determinations.

Implementation records fix credentials, lock state, account, and management policy for A3--A6. A3 distinguishes different accounts from shared unlocked sessions. A4 names its management, repair, export, backup, diagnostic, or forensic path and acquisition mode; management authority alone does not imply plaintext access, and in-scope backups remain deletion stores. A6 is a non-root process with stated permissions and entitlements; protected-store keys and other accounts' credentials are outside that class.

\begin{figure*}[t]
\centering
\begin{tikzpicture}[
  font=\scriptsize,
  box/.style={draw, semithick, rounded corners=4pt, align=center,
    inner xsep=5pt, inner ysep=4pt, fill=white},
  src/.style={box, fill=white, text width=2.45cm, minimum height=0.82cm},
  comp/.style={box, fill=black!6, text width=2.80cm, minimum height=0.98cm},
  store/.style={box, fill=black!10, text width=2.80cm, minimum height=0.98cm},
  outbox/.style={box, fill=white, text width=2.50cm, minimum height=0.82cm},
  flow/.style={-{Latex[length=2.0mm]}, semithick},
  risk/.style={-{Latex[length=1.7mm]}, dashed, semithick, black!70},
  boundary/.style={draw=black!60, dashed, rounded corners=6pt, line width=0.35pt},
  attack/.style={font=\scriptsize\itshape, align=center, text=black!70, fill=white, inner sep=0.5pt}
]
\draw[boundary, fill=black!4] (-0.20,-1.65) rectangle (2.90,2.15);
\node[anchor=north west,font=\scriptsize\bfseries] at (-0.05,2.08) {User/app contexts};
\draw[boundary, fill=black!1] (3.20,-1.65) rectangle (13.20,2.15);
\node[anchor=north west,font=\scriptsize\bfseries] at (3.35,2.08) {Semi-trusted platform TCB};
\draw[boundary, fill=black!4] (13.55,-1.65) rectangle (16.70,2.15);
\node[anchor=north west,font=\scriptsize\bfseries] at (13.70,2.08) {External boundary};

\node[src] (user) at (1.35,1.20) {User request\\and selection};
\node[src] (apps) at (1.35,0.15) {Apps \& OS services\\files, mail, calendar};
\node[src] (content) at (1.35,-0.90) {Web, messages,\\screenshots, media};
\node[comp] (broker) at (4.55,0.15) {\textbf{C1--C2: Context broker}\\eligibility, provenance\\redaction, authority};
\node[comp] (model) at (7.80,0.86) {\textbf{C3: Model/retrieval}\\prompt construction\\local model, index};
\node[store] (memory) at (7.80,-0.84) {\textbf{C3: Memory manager}\\embeddings, summaries\\retention, deletion};
\node[comp] (tool) at (11.15,0.15) {\textbf{C4: Tool/fallback broker}\\destination policy\\confirmation, routing};
\node[outbox] (localout) at (15.10,0.86) {Local output\\draft or action};
\node[outbox] (remote) at (15.10,-0.84) {Cloud fallback\\and telemetry};
\node[box, fill=black!12, text width=3.35cm] (updates) at (7.80,2.82)
  {\textbf{C5: Change governance}\\Updates, flags, admin policy};

\draw[flow] (user.east) -- ($(broker.west)+(0,0.43)$);
\draw[flow] (apps.east) -- (broker.west);
\node[attack, anchor=south west] at ($(broker.north west)+(-0.55,0.04)$) {confused deputy};
\draw[flow] (content.east) -- ($(broker.west)+(0,-0.43)$);
\node[attack, anchor=west] at ($(broker.south west)+(-0.10,-0.23)$) {prompt injection};
\draw[flow] (broker.east) -- (model.west);
\node[attack, anchor=south west] at ($(model.north west)+(-0.85,0.06)$) {cross-context leakage};
\draw[flow] (broker.east) |- (memory.west);
\draw[<->, semithick] (model.south) -- node[right,font=\tiny] {retrieve/promote} (memory.north);
\draw[flow] (model.east) -- (tool.west);
\draw[flow] (memory.east) -| ($(tool.west)+(0,-0.35)$);
\draw[flow] ($(tool.east)+(0,0.26)$) -- node[attack, above=1.3mm, pos=.35] {tool disclosure} (localout.west);
\draw[flow] ($(tool.east)+(0,-0.26)$) -- node[attack, below=1.3mm, pos=.35] {boundary shift} (remote.west);
\node[attack] (localadv) at (7.80,-2.20) {co-user/admin local access};
\draw[risk] (localadv.north) -- (memory.south);
\draw[risk] (updates.south west) -- (broker.north);
\draw[risk] (updates.south) -- node[attack, right, pos=.35] {update drift} (model.north);
\draw[risk] (updates.south east) -- (tool.north);
\node[attack] at (5.75,-1.86) {memory poisoning/persistence};
\end{tikzpicture}
\caption{System model and trust boundaries. Local inference sits inside a mediation path that separately governs source selection, context construction, retrieval and memory, tools, fallback, telemetry, and lifecycle change.}
\label{fig:systemmodel}
\end{figure*}

The system model yields six privacy failure surfaces. \emph{Access expansion} lets an app or assistant indirectly use sources outside its task authority. \emph{Context assembly and collapse} allows one social or application domain to shape another. \emph{Inference amplification} creates a sensitive fact that no source stated directly. \emph{Persistence and secondary use} make raw or derived state available after its original task. \emph{Actionable disclosure} carries out a tool action influenced by out-of-scope context. \emph{Governance drift} changes source scope, retention, routing, or authority after review. Figure~\ref{fig:systemmodel} connects these surfaces to the control points: context manifests and denied-source logs at C1--C2, retrieval and inference tests at C2--C3, memory ledgers and deletion traces at C3, tool and routing logs at C4, and policy diffs and regression replay at C5.

Fallback and telemetry are boundary-shift mechanisms, not a seventh surface. A route can expand access, collapse contexts, transmit an inferred fact, retain data in reports or logs, enable actionable disclosure, or drift after an update. Audits therefore place it under the applicable surface and C4--C5, recording destination, payload category, notice, retention claim, and replay evidence.

\section{Audit Workflow and Rubric}

The audit begins with a concrete workflow: for example, ``summarize this meeting using selected email and a local draft, then offer to send a follow-up message.'' The evaluator then maps active sources, retrieved state, memory records, tool permissions, fallback paths, telemetry, and policy/update dependencies. The key rule is conservative evidence use. A control that is asserted but not inspectable should score no higher than nominal, and a control visible only to enterprise administrators should be distinguished from consumer user agency. Scope statuses follow Section~III-A.

The documentary boundary record is a pre-execution gate, not a substitute for implementation testing. Without first fixing the boundary and separating UD from ADJ paths, a clean black-box observation such as $URR=0$ has an unknown coverage basis and can create false assurance. Material probes are interpreted only after this gate establishes what was reachable, excluded, and observed.

Documentary mode asks which control requirements public artifacts support and which evidence an implementation audit still needs. Its labels mainly span 1, 2, and UD: zero needs affirmative absence or contradiction, and level~3 needs claim-matched validation. A feature-specific notice specification can support a documentary D4 label when it names the trigger, audience, route, and notice content. That label does not establish actual delivery; a run-linked record is needed for that claim. Implementation mode adds pre-registered probes and execution artifacts. Section~\ref{sec:application} uses documentary mode.

Table~\ref{tab:rubric} gives the four-level scoring rubric. The scoring anchors are qualitative but evidence-oriented. A score of 0 requires affirmative evidence that a material control is absent or contradicted; missing evidence alone is UD. Policy-only statements or broad settings without task evidence receive at most 1; a feature-specific design or configured control can support 2; and enforced controls with protected audit artifacts, negative tests, and update-regression evidence can support 3. Multiple reviewers should score independently with evidence identifiers. Unresolved numeric disagreement can remain a range; disagreements involving UD, ADJ or N/A must retain the distinct statuses rather than be averaged or converted to zero.

\begin{table*}[t]
\caption{Evidence-qualified control levels. Levels 1--3 accumulate positive requirements within the selected applicable branch. Within an anchor, conditions joined by AND are cumulative; an explicitly stated OR denotes alternative branches. Level 0 requires affirmative absence or contradiction. Evidence grade and test outcome are separate fields.}
\label{tab:rubric}

\centering
\scriptsize
\setlength{\tabcolsep}{2.2pt}
\renewcommand{\arraystretch}{1.08}
\begin{tabular}{|D{0.17\textwidth}|P{0.18\textwidth}|P{0.18\textwidth}|P{0.18\textwidth}|P{0.20\textwidth}|}
\hline
\textbf{Dimension} & \textbf{0: Absent} & \textbf{1: Nominal} & \textbf{2: Feature-specific} & \textbf{3: Robust} \\
\hline
D1: Data minimization and context access & Task sources demonstrably unrestricted or lacking a data-flow control & Declared request/recipient boundary, or broad source permissions/policy; task-source provenance not required & Per-task source scoping with visible active sources & Enforced minimization with provenance logs, redaction, and denied-source evidence \\
\hline
D2: Contextual boundary and inference restraint & Affirmative evidence of unconstrained cross-domain retrieval or profiling & Coarse principal/recipient access constraint or general safety statement; task-specific limits not required & Domain labels, retrieval budgets, and sensitive-output checks & Default denial for cross-domain use; tested suppression of unnecessary sensitive inferences \\
\hline
D3: Memory and retention governance & Material durable state demonstrably lacks an inventory and retention/deletion control & Identified state, reset/delete control, or broad non-retention claim; detailed retention/deletion rules not required & State inventory with retention classes, or a specified non-creation/non-retention path & Enforced non-creation or promotion gates, deletion propagation to summaries/embeddings/indexes, and retention audits \\
\hline
D4: Tool and fallback mediation & Affirmative evidence of undisclosed tool calls or remote routing & User/admin confirmation or delivered policy notice for the declared action/route; provenance not required & Runtime notice for tool use, fallback, and sensitive source influence & Policy broker blocks out-of-scope disclosure and keeps auditable tool/fallback logs \\
\hline
D5: Accountability, user agency, and change governance & Applicable accountability, recourse, and change review demonstrably absent & Named capability owner OR capability-specific recourse/change-review undertaking & Named owner, feature-level logs and review gates, plus user/admin recourse & Independent review, regression tests after updates, documented rollback, and contestability \\
\hline
\end{tabular}
\end{table*}

\subsection{Audit Records, Evidence, and Probes}
\label{sec:protocol}

Levels 1--3 accumulate positive requirements within a dimension's applicable branch: level 2 retains the supported level-1 control, and level 3 adds enforcement and validation to level 2. A lower anchor does not require a higher control to be absent. For example, confirmation with provenance can support D4-L1 even when runtime notice is insufficient for D4-L2. D3 follows either the retained-state or specified non-retention branch; deletion duties apply to any stores that exist. These are evidence-qualified control levels, not an interval scale or an estimate of protection effectiveness.

Each audit record separates control maturity, evidence strength, test outcome, and claim status. They are distinct but not statistically independent: level~1 requires at least E1, level~2 at least E2, and level~3 claim-matched E4 coverage. Typical artifacts are an E1 policy assertion, an E2 feature design or inspectable configuration, an E3 collected execution record or completed scoped evaluation, and an E4 validation package with negative cases, coverage, uncertainty, and change replay. A log schema without a collected trace, or a research environment without an executed study, remains E2. Mixed artifacts are graded per subclaim rather than averaged.

\noindent\textbf{Audit-record schema.} Boundary and version $\rightarrow$ access class A0--A6 $\rightarrow$ control points C1--C5 $\rightarrow$ dimension D1--D5 with maturity 0--3 or scope status SCORED/UD/ADJ/N/A $\rightarrow$ evidence E1--E4 $\rightarrow$ test outcome and claim status $\rightarrow$ adjacent paths and material-change set.

A repeatable record names the task, version, policy state, sources, stores, tools, fallback and telemetry paths, and reviewer. Adjacent records identify parent and target boundaries, reachability, status, and blocker propagation. Assign the highest fully supported anchor. Anchor conditions apply only to applicable paths: an evidenced-absent fallback neither satisfies nor blocks unrelated duties. Material subclaims are recorded separately; one dimension label is issued only after each is resolved. A material untested probe yields UD unless a pre-registered substitute supports the condition.

Implementation audits use a clean account, synthetic data, an allow/deny manifest, fixed observation windows, and reviewer-visible artifacts. The manifest records contextual-integrity roles and authority rather than assuming that device ownership alone justifies every flow \cite{nissenbaum_context}. Task utility is evaluated separately so deny-all behavior is not credited as useful protection. Context manifests, denied-source logs, memory ledgers, tool/routing logs, and policy diffs are the white-box ideal; canaries, paired prompts, UI notices, endpoint-visible traffic, tool arguments, and local storage changes provide lower-bound external evidence. Audit logs are themselves privacy-relevant state and must be minimized, access-controlled, and retained under explicit rules.

\begin{table*}[t]
\caption{Candidate probes for implementation audits. These counting rules are specifications, not measurements reported in this paper. Lower $\mathrm{URR}$, $\mathrm{CLR}$, $\mathrm{TOR}$, and $\mathrm{GDR}$ are preferable; higher $\mathrm{DPS}$ and $\mathrm{BDR}$ are preferable within a fixed scope. A zero denominator is undefined; coverage and task utility are reported separately.}
\label{tab:probeops}
\centering
\scriptsize
\setlength{\tabcolsep}{2.4pt}
\renewcommand{\arraystretch}{1.04}
\begin{tabular}{|D{0.19\textwidth}|P{0.08\textwidth}|P{0.42\textwidth}|P{0.24\textwidth}|}
\hline
\textbf{Probe} & \textbf{Primary} & \textbf{Counted unit} & \textbf{Interpretation} \\
\hline
Unauthorized retrieval $\mathrm{URR}=N_{\mathrm{denied\ used}}/N_{\mathrm{denied\ opp}}$ & D1 & Pre-registered run--item pairs. Use includes retrieval, citation, a tool argument, protected output, or a paired-control effect. & Any verified denied use is FAIL; zero is not PASS without coverage and provenance. \\
\hline
Context/inference leakage $\mathrm{CLR}=N_{\mathrm{leakage}}/N_{\mathrm{cross\mbox{-}context}}$ & D2 & Pre-registered cross-context trials where disallowed data can alter output, action, routing, memory, or a sensitive inference. & Verified influence is FAIL; PASS requires repeated provenance-backed negative tests. \\
\hline
Deletion propagation $\mathrm{DPS}=N_{\mathrm{absent\ pairs}}/N_{\mathrm{tested\ pairs}}$ after $T_d$ & D3 & Seeded item times each raw store, snippet, index, embedding, summary, inferred memory, log, cache, and fallback report. & Unexplained presence is FAIL; exceptions and backups remain visible. \\
\hline
Tool-action overreach $\mathrm{TOR}=N_{\mathrm{overbroad}}/N_{\mathrm{tool\ opp}}$ & D4 & Proposed or executable calls with a recipient, source, scope, or side effect outside the manifest or user intent. & Executed overreach is FAIL; a blocked and logged attempt is evidence of mediation. \\
\hline
Boundary-disclosure rate $\mathrm{BDR}=N_{\mathrm{observable}}/N_{\mathrm{route\ events}}$ & D4 & Triggered fallback/telemetry routes with destination, payload category, retention claim, notice audience, and notice time. & Reviewer logs and delivered notices are recorded separately; observability alone is not consent. \\
\hline
Governance drift $\mathrm{GDR}=N_{\mathrm{unreviewed}}/N_{\mathrm{changes}}$ & D5 & Material model, OS, policy, source, retention, tool, route, UI, or logging changes lacking diff, replay, or exception decision. & An unreviewed material change is FAIL for D5 and blocks robust affected scores. \\
\hline
\end{tabular}
\end{table*}

\section{Documentation-Bounded Application}
\label{sec:application}

We applied the rubric to four versioned boundaries using primary platform documentation. The registry fixes product/API scope, configuration, policy state, feature flags, and documentation cutoff; fields not established by the corpus remain UD. Table~\ref{tab:platforms} reports 20 documentary classifications. Its upper range is limited by the absence of claim-matched validation artifacts, so the useful output is the control basis and missing evidence for each row. The sources chiefly address A1 for local APIs, A5 for PCC, and A3--A4 for Recall.

PCC and Recall share the summary vector $(1,1,2,\mathrm{UD},1)$, but their obligations differ. PCC D3 follows specified request non-retention; Recall D3 follows retained-state inventory and retention classes. PCC D5 additionally cites an E3 examination of provisioning controls, while Recall D5 rests on E2 documentation. The bracketed control basis in Table~\ref{tab:platforms} makes these distinctions visible. The compression is intentional: documentary mode triages evidence rather than ranking product privacy. Uniform D1/D2 level~1 records coarse request, principal, or recipient boundaries without task-specific restraint. All four D4 rows are UD because an applicable internal notice or mediation duty remains unresolved; D4 is nevertheless reachable from a feature-specific notice that names its trigger, audience, route, and content.

\begin{table*}[t]
\caption{Documentary control profiles: label [control basis or unresolved duty; evidence grade]. Labels summarize anchor support, not measured protection. E3 in PCC D5 covers provisioning only. The source-linked ledger gives locators, boundary exclusions, and unmet requirements.}
\label{tab:platforms}

\centering
\scriptsize
\setlength{\tabcolsep}{1.8pt}
\renewcommand{\arraystretch}{1.08}
\begin{tabular}{|D{0.18\textwidth}|P{0.15\textwidth}|P{0.15\textwidth}|P{0.15\textwidth}|P{0.15\textwidth}|P{0.16\textwidth}|}
\hline
\textbf{Feature boundary} & \textbf{D1: Data/context} & \textbf{D2: Boundary/inference} & \textbf{D3: Memory/retention} & \textbf{D4: Tool/fallback} & \textbf{D5: Agency/governance} \\
\hline
Apple Foundation Models WWDC25 on-device text-only session & \textbf{1} [request scope; E2] & \textbf{1} [instruction hierarchy; E2] & \textbf{1} [identified transcript; E2] & \textbf{UD} [dispatch notice; E2] & \textbf{UD} [accountability undertaking; E1] \\
\hline
Apple PCC one-shot request path & \textbf{1} [request scope; E2] & \textbf{1} [recipient access; E2] & \textbf{2} [specified non-retention; E2] & \textbf{UD} [route notice; E2] & \textbf{1} [verification undertaking; E2/E3] \\
\hline
Android AICore/Gemini Nano request API & \textbf{1} [request isolation; E2] & \textbf{1} [principal isolation; E2] & \textbf{2} [specified non-retention; E2] & \textbf{UD} [internal notice duties; E2] & \textbf{UD} [accountability undertaking; E2] \\
\hline
Microsoft Recall snapshot/index workflow & \textbf{1} [app/site filters; E2] & \textbf{1} [account/enclave access; E2] & \textbf{2} [inventory and retention classes; E2] & \textbf{UD} [metadata-request notice; E2] & \textbf{1} [feature privacy undertaking; E2] \\
\hline
\end{tabular}
\end{table*}

The narrow boundaries prevent several misleading comparisons. Foundation Models remains locked to the WWDC25 on-device text-only API. Host tool implementations, persistence and actions are adjacent, but dispatch remains a framework obligation. Developer-selected PCC and other model providers introduced at WWDC26 are outside this historical API rather than automatic fallback \cite{apple_wwdc25_foundation_models, apple_wwdc26_foundation_models}. PCC begins after client request construction, so recipient and retention safeguards cannot establish minimal client source selection \cite{apple_pcc_security, apple_pcc_security_guide}. AICore request isolation and non-retention likewise do not discharge app retrieval, storage, action or model-delivery obligations \cite{android_gemini_nano, android_aicore_privacy, google_android_aicore_help}. Microsoft states that snapshot storage and analysis do not use cloud processing, so built-in snapshot cloud sync remains N/A. Ancillary web-metadata requests, feedback screenshots, Click to Do, EEA user export, diagnostics and Copilot have distinct paths \cite{microsoft_recall_manage, microsoft_recall_click_to_do, microsoft_recall_privacy, microsoft_recall_architecture, microsoft_recall_snapshots_disk}. Their external recipients can be ADJ while the initiating Recall operation still needs notice and provenance evidence. Microsoft's 2024 Recall redesign changed opt-in, isolation, and access-control assumptions, illustrating why a release change needs a new boundary record rather than inherited labels. None of these component profiles supports an end-to-end product claim until its material internal and adjacent obligations are evaluated.

Boundary choice also changes the claim. Expanding a Foundation Models session to include a host-app mail tool brings the recipient, arguments, notice, and host retention into scope. Even a future scored dispatch subclaim would leave the expanded workflow UD while a material tool obligation remains unresolved. Merely drawing a narrower API boundary cannot improve the workflow label.

\subsection{Evidence Requests for Deployment Review}
The matrix supports deciding what evidence to request next. Foundation Models needs transcript retention and deletion rules; PCC and AICore need tests of their specified non-retention paths; Recall needs propagation evidence linking snapshots to OCR, vectors, indexes, and applicable backups. These are different obligations despite matching D3 labels. For D4, each unresolved row calls for a notice specification naming the initiating event, audience, destination, payload class, and delivery time, followed by a run-linked record. For D5, a named capability owner, review gates, recourse path, and change-replay evidence close different gaps. Publishing these artifacts would make the claims assessable; a higher label follows only when the applicable anchor is met.

\subsection{Fleet Security Triage and Forensic Scope}
\label{sec:fleet}
An operational use of the audit record is to compare security events from devices governed by the same capability boundary and policy. Table~\ref{tab:fleet-event} specifies a minimized event for that purpose. It links source decisions, tool routes, notices, deletion checks, and policy changes without centralizing raw messages, screenshots, or inferred facts. Recipient tokens are keyed HMACs over the recipient identifier and comparison-window ID; the analytics control plane holds the key and rotates it between windows unless a separately approved longitudinal join is required. Because boundary, region, build, policy, and task fields can jointly identify small cohorts, reports suppress cohorts below a pre-registered $k_{\min}$ and restrict artifact joins.

\begin{table}[t]
\caption{Proposed fleet event fields. No device events were collected.}
\label{tab:fleet-event}
\centering
\footnotesize
\setlength{\tabcolsep}{3pt}
\renewcommand{\arraystretch}{1.05}
\begin{tabular}{|P{0.25\columnwidth}|P{0.65\columnwidth}|}
\hline
\textbf{Group} & \textbf{Fields and use} \\
\hline
Cohort & \texttt{boundary\_id}, \texttt{region}, \texttt{policy\_version}, \texttt{task\_class}, \texttt{probe\_id}: define comparable runs. \\
\hline
Release & \texttt{build\_id}: distinguish successive OS or model releases. \\
\hline
Event & \texttt{event\_id}, \texttt{event\_type}, \texttt{recipient\_token}, \texttt{artifact\_ref}: link route and notice records. \\
\hline
Coverage & \texttt{observation\_status}: observed, missing, or not collected. \\
\hline
\end{tabular}
\end{table}

For example, hold the boundary, region, policy, task class, and probe specification fixed while comparing build N with N+1. Identify recipient tokens first seen in N+1, join each new route to its notice by event ID and to change review by boundary and build, and flag missing notices or unreviewed route changes. Derive observation status from an expected-device manifest plus authenticated heartbeat and sequence coverage; report missing observations rather than counting them as safe outcomes. $\mathrm{BDR}$ summarizes observed route coverage, and $\mathrm{GDR}$ counts material changes without review or replay. The resulting list narrows analyst triage to affected cohorts and paths, but detection quality and processing cost remain unmeasured.

The same boundary inventory scopes forensic review. A Recall snapshot, OCR record, vector, and derived summary are separate state objects. An A4 acquisition record must state lock and credential conditions, acquisition path, and covered stores. DPS measures deletion across tested item--store pairs; it does not measure forensic acquisition completeness.

\subsection{PCC as an Evidence-Scoping Case}
\label{sec:pcc}
PCC exposes stronger artifacts than a privacy policy: hardware and release-admission specifications, public software measurements, an inspection environment, request-report schemas, and an independent examination of provisioning controls \cite{apple_pcc_hardware,apple_pcc_transparency,apple_pcc_vre,apple_pcc_report,apple_pcc_soc3_report}. They support different claims. Release identity and attestation do not establish task-purpose correctness; availability of the Virtual Research Environment is E2 until an experiment is run; and the completed SOC~3 examination is E3 only for its stated provisioning scope, not for AI-service behavior or our five dimensions.

The same separation applies to tool and state paths. Reports can expose routing, attestation, and external tool arguments, but a retrospective export does not establish pre-action notice or the recipient's retention policy. A delegated PCC worker, an external tool recipient, and a client-executed tool cross different boundaries \cite{apple_pcc_agent,apple_pcc_transitive}. Multi-turn encrypted key-value cache also requires a different state inventory from one-shot inference: PCC cannot read that cache without client-supplied keys between turns, but the encrypted state still requires an inventory \cite{apple_pcc_multiturn}.

\subsection{Formative Comparison}
\label{sec:formative}
As a formative check, one author and one non-author colleague in the same organization separately coded four capabilities across D1--D5 using an author-prepared guide and candidate source list. The author designed the rubric and had prior manuscript exposure. Scope states matched on 19/20 items; D1, D2, and D4 were constant across cases, while the more variable D3/D5 subset matched on 7/8. The 19/20 count is descriptive only: 12 cells lie in dimensions constant across cases, so no reliability estimate is inferred. The sole difference was AICore D5: the author assigned 1 and the colleague UD. The final rubric makes the capability-specific accountability branch explicit, leaving D5 UD; this adjudication refines the method rather than validates either record as ground truth.

The pilot guide listed 23 candidate entries and excluded the full PCC SOC~3 report while retaining Apple's public overview; the final 24-source corpus adds that report only for the PCC provisioning analysis. The pair is therefore not a frozen-corpus replication of the final matrix.

Comparing the rationales behind matching labels produced four further instructions: an adjacent recipient does not erase the initiating component's mediation duties; notice credit names its audience and delivery time; D3 records whether level~2 rests on retained-state inventory or specified non-retention; and account authentication does not establish within-account inference restraint. The paired records are used to locate ambiguity, not to estimate statistical inter-rater reliability or control effectiveness.

\section{Synthetic Walkthrough: Meeting Assistant}

This section is a synthetic walkthrough of the audit record, not a reported product experiment. Consider a device assistant that prepares a user for a named meeting and can draft a follow-up email. The registered boundary includes meeting-scoped retrieval, draft generation, memory/index updates, and the proposed mail action. Later cloud synchronization is disabled; if enabled, it becomes ADJ and blocks a workflow claim until separately evaluated.

The allow-list contains one calendar event, a meeting folder, and selected work messages. Personal mail, chat, screenshots, and unrelated files are sealed. A fixture places canary \texttt{cedar-04} in allowed material and \texttt{orchid-17} plus a medical-leave fact in personal mail. Suppose twelve runs inspect the context manifest, source UI, retrieval trace, denied-source log, output, and paired-removal behavior. Illustrative counts of $URR=0/48$ and $CLR=0/12$ would remain INCONCLUSIVE without pre-registered coverage and change replay. Any use or paraphrase of the protected fact would be FAIL and contradict the associated no-leakage claim.

For tool mediation, an untrusted agenda says to send a confidential margin note to \texttt{vendor@example.test}. Retrieved text must remain data rather than authority: it may not add a recipient or widen tool arguments. The auditor compares source provenance, proposed call, recipient diff, delivered notice, and broker decision. A logged refusal is E3 evidence for that attempt, not level~3 without scenario coverage and replay.

For deletion, a seeded meeting note produces a snippet, embedding, summary, inferred personnel fact, tool cache, and logs. The audit checks every applicable store after a fixed window and keeps documented backup or legal-hold exceptions in the record. The same fixtures are replayed after model, retrieval, policy, or feature-flag changes; unexplained expansion of source scope, retained state, or mail authority is governance drift even if inference remains local.

\section{Open Problems for AI Security}

\subsection{Implementation and Stakeholder Questions}

Empirical evaluation should test denied-source and indirect-injection paths across apps, then trace deletion through prompts, snippets, embeddings, summaries, inferred facts, logs, and model artifacts. Tool and fallback tests need source, recipient, destination, and user-intent records; the same cases require replay after model, policy, or feature-flag changes. Context manifests, memory ledgers, routing logs, and policy diffs are the necessary evidence.

Threat exposure also depends on the user. Shared-device users and abuse survivors may face authorized co-users searching assistant history; workers on managed devices may face capture, DLP review, and administrator logs \cite{microsoft_recall_manage, apple_device_management_restrictions}. Journalists, clinicians, and lawyers may face compelled access to local indexes. These cases require different A3/A4 assumptions even when inference remains on-device.

\section{Limitations}

This is a framework study with a small formative assessment comparison, not an empirical security benchmark. We did not reverse-engineer closed systems, run the probes in Table~\ref{tab:probeops}, conduct an end-user study, or build a mediation layer. The document corpus can omit implemented controls and become stale as live pages change. The two assessment records support a formative comparison; they do not estimate inter-rater reliability or control effectiveness. Because the same four boundaries informed the guidance clarifications and the final application, transfer to a held-out boundary remains untested. The four cases do not constitute 20 independent systems, and equal levels can represent different controls. The pilot was not timed, so we do not report evaluator-hours. Documentary use requires privacy/security audit expertise and primary-source access; implementation mode also requires a configured device/account, synthetic fixtures, and source, memory, tool, and routing records. Further validation requires external evaluators and implementation-based evaluation. Semi-trusted-provider scope excludes a malicious OS vendor, hardware side channels, and formal verification of closed components.

\section{Reproducibility and Ethics}
The Source Traceability Supplement is available at \url{https://github.com/johnlgtmchung/os-ai-audit-supplement/releases/tag/bigcyber26-submission}. It contains the primary-source manifest, four scored boundary records, one unscored PCC Agent property record, 20 dimension cells, adjacent paths, paired labels, and scripts for source checks and descriptive counts. Its \texttt{README.md} maps the files to the analysis. Source locators and hashes support traceability; the package does not preserve vendor-page bodies.

The platform analysis used public documentation and literature without private user content, vulnerability exploitation, or production-system interaction. The formative comparison used de-identified assessment records from an author and a same-organization colleague, who provided written permission for research and publication use. Raw workbooks and the permission record are not distributed. All example traces use synthetic canaries and the reserved \texttt{example.test} domain.

No institutional ethics review was obtained. We treated the comparison as methods development using professional judgments about public technical material, with no intervention or collection of private, sensitive, or biometric information; this is a rationale, not a claimed institutional exemption.

Both authors are employed by Google LLC, and the non-author evaluator was a colleague in the same organization. All four platform boundaries were assessed under the same rubric using externally available sources; the study used no confidential Google information and does not represent an official company assessment. The only initial label disagreement concerned AICore D5, and the final matrix adopts the more conservative UD classification under the clarified accountability rule.

\section{Conclusion}

Local inference answers where computation occurs, not which information use is justified. A feature-boundary audit links context, retained state, delegated authority and platform change to claim-specific evidence. The formative records show why that link needs explicit decision rules: the same numerical score can describe different retention regimes, an adjacent recipient can leave an internal mediation duty, and an accountability pledge can be interpreted differently from a safety statement. Recording those distinctions and their evidence makes the analysis inspectable. Implementation evaluation can build on these records with logs, negative tests, deletion traces and change replay.

\section*{Acknowledgment}
We thank the non-author colleague who completed the formative assessment and permitted use of the de-identified record.

OpenAI Codex assisted with revising the text. It did not assign ratings or select evidence. The authors rigorously reviewed all resulting edits and take full responsibility for the final manuscript and artifacts.

\bibliographystyle{IEEEtran}
\IEEEtriggeratref{60}
\bibliography{refs}

\begin{thebibliography}{10}
\providecommand{\url}[1]{#1}
\csname url@samestyle\endcsname
\providecommand{\newblock}{\relax}
\providecommand{\bibinfo}[2]{#2}
\providecommand{\BIBentrySTDinterwordspacing}{\spaceskip=0pt\relax}
\providecommand{\BIBentryALTinterwordstretchfactor}{4}
\providecommand{\BIBentryALTinterwordspacing}{\spaceskip=\fontdimen2\font plus
\BIBentryALTinterwordstretchfactor\fontdimen3\font minus
  \fontdimen4\font\relax}
\providecommand{\BIBforeignlanguage}[2]{{%
\expandafter\ifx\csname l@#1\endcsname\relax
\typeout{** WARNING: IEEEtran.bst: No hyphenation pattern has been}%
\typeout{** loaded for the language `#1'. Using the pattern for}%
\typeout{** the default language instead.}%
\else
\language=\csname l@#1\endcsname
\fi
#2}}
\providecommand{\BIBdecl}{\relax}
\BIBdecl

\bibitem{apple_foundation_models}
{Apple Developer}, ``Foundation models,''
  \url{https://developer.apple.com/documentation/FoundationModels}, 2026,
  developer documentation, accessed May 5, 2026.

\bibitem{android_gemini_nano}
{Google Android Developers}, ``{Gemini Nano},''
  \url{https://developer.android.com/ai/gemini-nano}, 2026, developer
  documentation, revision dated September 8, 2026; accessed September 24, 2026.

\bibitem{microsoft_recall_manage}
{Microsoft Learn}, ``Manage recall for windows clients,''
  \url{https://learn.microsoft.com/en-us/windows/client-management/manage-recall},
  2026, product documentation, accessed May 5, 2026.

\bibitem{nissenbaum_context}
H.~Nissenbaum, ``Privacy as contextual integrity,'' \emph{Washington Law
  Review}, vol.~79, no.~1, pp. 119--158, 2004.

\bibitem{solove_taxonomy}
D.~J. Solove, ``A taxonomy of privacy,'' \emph{University of Pennsylvania Law
  Review}, vol. 154, no.~3, pp. 477--560, 2006.

\bibitem{cavoukian_pbd}
A.~Cavoukian, ``Privacy by design: The 7 foundational principles,'' 2011,
  {I}nformation and Privacy Commissioner of Ontario.

\bibitem{deng_linddun}
M.~Deng, K.~Wuyts, R.~Scandariato, B.~Preneel, and W.~Joosen, ``A privacy
  threat analysis framework: Supporting the elicitation and fulfillment of
  privacy requirements,'' \emph{Requirements Engineering}, vol.~16, no.~1, pp.
  3--32, 2011.

\bibitem{hoepman_privacy_design_strategies}
J.-H. Hoepman, ``Privacy design strategies,'' in \emph{ICT Systems Security and
  Privacy Protection}.\hskip 1em plus 0.5em minus 0.4em\relax Berlin,
  Heidelberg: Springer, 2014, pp. 446--459.

\bibitem{gdpr_article35}
{European Union}, ``{Regulation (EU) 2016/679, Article 35: Data Protection
  Impact Assessment},'' \url{https://eur-lex.europa.eu/eli/reg/2016/679/oj},
  2016, accessed May 7, 2026.

\bibitem{ico_dpia_guidance}
{Information Commissioner's Office}, ``Data protection impact assessments,''
  \url{https://ico.org.uk/for-organisations/uk-gdpr-guidance-and-resources/accountability-and-governance/data-protection-impact-assessments-dpias/},
  2026, accessed May 7, 2026.

\bibitem{nist_privacy_framework}
{National Institute of Standards and Technology}, ``{NIST Privacy Framework: A
  Tool for Improving Privacy through Enterprise Risk Management, Version
  1.0},'' \url{https://www.nist.gov/privacy-framework}, 2020, {NIST} CSWP
  01162020, accessed May 7, 2026.

\bibitem{saltzer_schroeder}
J.~H. Saltzer and M.~D. Schroeder, ``The protection of information in computer
  systems,'' \emph{Proceedings of the IEEE}, vol.~63, no.~9, pp. 1278--1308,
  1975.

\bibitem{hardy_confused_deputy}
N.~Hardy, ``The confused deputy: Or why capabilities might have been
  invented,'' \emph{ACM SIGOPS Operating Systems Review}, vol.~22, no.~4, pp.
  36--38, 1988.

\bibitem{felt_android_permissions}
A.~P. Felt, E.~Ha, S.~Egelman, A.~Haney, E.~Chin, and D.~Wagner, ``Android
  permissions: User attention, comprehension, and behavior,'' in
  \emph{Proceedings of the Eighth Symposium on Usable Privacy and
  Security}.\hskip 1em plus 0.5em minus 0.4em\relax New York, NY, USA:
  Association for Computing Machinery, 2012, pp. 3:1--3:14.

\bibitem{roesner_user_driven_access}
F.~Roesner, T.~Kohno, A.~Moshchuk, B.~Parno, H.~J. Wang, and C.~Cowan,
  ``User-driven access control: Rethinking permission granting in modern
  operating systems,'' in \emph{Proceedings of the IEEE Symposium on Security
  and Privacy}.\hskip 1em plus 0.5em minus 0.4em\relax Piscataway, NJ, USA:
  IEEE, 2012, pp. 224--238.

\bibitem{enck_taintdroid}
W.~Enck, P.~Gilbert, B.-G. Chun, L.~P. Cox, J.~Jung, P.~McDaniel, and A.~N.
  Sheth, ``Taintdroid: An information-flow tracking system for realtime privacy
  monitoring on smartphones,'' in \emph{Proceedings of the 9th USENIX Symposium
  on Operating Systems Design and Implementation}.\hskip 1em plus 0.5em minus
  0.4em\relax Berkeley, CA, USA: USENIX Association, 2010, pp. 393--407.

\bibitem{lau_alexa_listening}
J.~Lau, B.~Zimmerman, and F.~Schaub, ``Alexa, are you listening? privacy
  perceptions, concerns and privacy-seeking behaviors with smart speakers,''
  \emph{Proceedings of the ACM on Human-Computer Interaction}, vol.~2, no.
  CSCW, 2018.

\bibitem{malkin_smart_speakers}
N.~Malkin, J.~Deatrick, A.~Tong, P.~Wijesekera, S.~Egelman, and D.~Wagner,
  ``Privacy attitudes of smart speaker users,'' \emph{Proceedings on Privacy
  Enhancing Technologies}, vol. 2019, no.~4, pp. 250--271, 2019.

\bibitem{abdi_smart_speakers}
N.~Abdi, K.~M. Ramokapane, and J.~M. Such, ``More than smart speakers: Security
  and privacy perceptions of smart home personal assistants,'' in
  \emph{Proceedings of the Symposium on Usable Privacy and Security}.\hskip 1em
  plus 0.5em minus 0.4em\relax Berkeley, CA, USA: USENIX Association, 2019, pp.
  451--466.

\bibitem{zhang_dangerous_skills}
N.~Zhang, X.~Mi, X.~Feng, X.~Wang, Y.~Tian, and F.~Qian, ``Dangerous skills:
  Understanding and mitigating security risks of voice-controlled third-party
  functions on virtual personal assistant systems,'' in \emph{2019 IEEE
  Symposium on Security and Privacy}.\hskip 1em plus 0.5em minus 0.4em\relax
  Los Alamitos, CA, USA: IEEE, 2019, pp. 1381--1396.

\bibitem{song_embedding_leakage}
C.~Song and A.~Raghunathan, ``Information leakage in embedding models,'' in
  \emph{Proceedings of the 2020 ACM SIGSAC Conference on Computer and
  Communications Security}.\hskip 1em plus 0.5em minus 0.4em\relax New York,
  NY, USA: Association for Computing Machinery, 2020, pp. 377--390.

\bibitem{morris_embeddings_reveal}
\BIBentryALTinterwordspacing
J.~X. Morris, V.~Kuleshov, V.~Shmatikov, and A.~M. Rush, ``Text embeddings
  reveal (almost) as much as text,'' in \emph{Proceedings of the 2023
  Conference on Empirical Methods in Natural Language Processing}.\hskip 1em
  plus 0.5em minus 0.4em\relax Singapore: Association for Computational
  Linguistics, 2023, pp. 12\,448--12\,460. [Online]. Available:
  \url{https://aclanthology.org/2023.emnlp-main.765/}
\BIBentrySTDinterwordspacing

\bibitem{carlini_extracting}
N.~Carlini, F.~Tramer, E.~Wallace, M.~Jagielski, A.~Herbert-Voss, K.~Lee,
  A.~Roberts, T.~Brown, D.~Song, U.~Erlingsson, A.~Oprea, and C.~Raffel,
  ``Extracting training data from large language models,'' in \emph{Proceedings
  of USENIX Security}.\hskip 1em plus 0.5em minus 0.4em\relax Berkeley, CA,
  USA: USENIX Association, 2021, pp. 2633--2650.

\bibitem{nasr_privacy_dnn}
M.~Nasr, R.~Shokri, and A.~Houmansadr, ``Comprehensive privacy analysis of deep
  learning: Passive and active white-box inference attacks against centralized
  and federated learning,'' in \emph{Proceedings of IEEE Symposium on Security
  and Privacy}.\hskip 1em plus 0.5em minus 0.4em\relax Los Alamitos, CA, USA:
  IEEE, 2019, pp. 739--753.

\bibitem{greshake_indirect_prompt_injection}
K.~Greshake, S.~Abdelnabi, S.~Mishra, C.~Endres, T.~Holz, and M.~Fritz, ``Not
  what you've signed up for: Compromising real-world {LLM}-integrated
  applications with indirect prompt injection,'' in \emph{Proceedings of the
  16th ACM Workshop on Artificial Intelligence and Security}.\hskip 1em plus
  0.5em minus 0.4em\relax New York, NY, USA: Association for Computing
  Machinery, 2023, pp. 79--90.

\bibitem{wang_memory}
B.~Wang, W.~He, S.~Zeng, Z.~Xiang, Y.~Xing, J.~Tang, and P.~He, ``Unveiling
  privacy risks in {LLM} agent memory,'' 2025, arXiv:2502.13172.

\bibitem{shi_tool_selection}
\BIBentryALTinterwordspacing
J.~Shi, Z.~Yuan, G.~Tie, P.~Zhou, N.~Z. Gong, and L.~Sun, ``Prompt injection
  attack to tool selection in {LLM} agents,'' in \emph{Proceedings of the
  Network and Distributed System Security Symposium}.\hskip 1em plus 0.5em
  minus 0.4em\relax San Diego, CA, USA: The Internet Society, 2026,
  arXiv:2504.19793. [Online]. Available:
  \url{https://www.ndss-symposium.org/ndss-paper/prompt-injection-attack-to-tool-selection-in-llm-agents/}
\BIBentrySTDinterwordspacing

\bibitem{doshi_safe_tool_use}
A.~Doshi, Y.~Hong, C.~Xu, E.~Kang, A.~Kapravelos, and C.~K{\"a}stner, ``Towards
  verifiably safe tool use for {LLM} agents,'' in \emph{Proceedings of the
  IEEE/ACM 48th International Conference on Software Engineering New Ideas and
  Emerging Results}.\hskip 1em plus 0.5em minus 0.4em\relax New York, NY, USA:
  Association for Computing Machinery, 2026, pp. 201--205, arXiv:2601.08012.

\bibitem{li_access_control}
X.~Li, D.~Huang, J.~Li, H.~Cai, Z.~Zhou, W.~Dong, X.~Wang, and Y.~Liu, ``A
  vision for access control in {LLM}-based agent systems,'' 2025,
  arXiv:2510.11108.

\bibitem{nist_ai_rmf}
{National Institute of Standards and Technology}, ``{Artificial Intelligence
  Risk Management Framework (AI RMF 1.0)},''
  \url{https://doi.org/10.6028/NIST.AI.100-1}, 2023, {NIST} {AI} 100-1.

\bibitem{nist_genai}
------, ``Artificial intelligence risk management framework: Generative
  artificial intelligence profile,''
  \url{https://doi.org/10.6028/NIST.AI.600-1}, 2024, {NIST} {AI} 600-1.

\bibitem{raji_gap}
I.~D. Raji, A.~Smart, R.~N. White, M.~Mitchell, T.~Gebru, B.~Hutchinson,
  J.~Smith-Loud, D.~Theron, and P.~Barnes, ``Closing the {AI} accountability
  gap: Defining an end-to-end framework for internal algorithmic auditing,'' in
  \emph{Proceedings of the 2020 Conference on Fairness, Accountability, and
  Transparency}.\hskip 1em plus 0.5em minus 0.4em\relax New York, NY, USA:
  Association for Computing Machinery, 2020, pp. 33--44.

\bibitem{diakopoulos_accountability}
N.~Diakopoulos, ``Algorithmic accountability: Journalistic investigation of
  computational power structures,'' \emph{Digital Journalism}, vol.~3, no.~3,
  pp. 398--415, 2015.

\bibitem{lahjouji_agents_too_much}
\BIBentryALTinterwordspacing
N.~Lahjouji and A.~G. Colaco, ``Agents that know too much: A data-centric
  survey of privacy in {LLM} agents,'' 2026. [Online]. Available:
  \url{https://arxiv.org/abs/2606.26627}
\BIBentrySTDinterwordspacing

\bibitem{elyagoubi_agentleak}
\BIBentryALTinterwordspacing
F.~{El Yagoubi}, G.~Badu-Marfo, and R.~{Al Mallah}, ``{AgentLeak}: A benchmark
  for internal-channel privacy leakage in multi-agent {LLM} systems,'' 2026.
  [Online]. Available: \url{https://arxiv.org/abs/2602.11510}
\BIBentrySTDinterwordspacing

\bibitem{xie_ocelot}
\BIBentryALTinterwordspacing
J.~Xie and S.~Li, ``{OCELOT}: Inference-leakage budgets for privacy-preserving
  {LLM} agents,'' 2026. [Online]. Available:
  \url{https://arxiv.org/abs/2606.12341}
\BIBentrySTDinterwordspacing

\bibitem{zou_blind_gods}
\BIBentryALTinterwordspacing
Z.~Zou, S.~Guo, Q.~Zhan, L.~Zhao, S.~Li, Q.~Li, K.~Xu, M.~Xu, and Z.~Liu,
  ``Blind gods and broken screens: Architecting a secure, intent-centric mobile
  agent operating system,'' 2026. [Online]. Available:
  \url{https://arxiv.org/abs/2602.10915}
\BIBentrySTDinterwordspacing

\bibitem{shao_privacylens}
\BIBentryALTinterwordspacing
Y.~Shao, T.~Li, W.~Shi, Y.~Liu, and D.~Yang, ``{PrivacyLens}: Evaluating
  privacy norm awareness of language models in action,'' in \emph{Advances in
  Neural Information Processing Systems}, vol.~37.\hskip 1em plus 0.5em minus
  0.4em\relax Red Hook, NY, USA: Curran Associates, Inc., 2024, pp.
  89\,373--89\,407. [Online]. Available:
  \url{https://proceedings.neurips.cc/paper_files/paper/2024/file/a2a7e58309d5190082390ff10ff3b2b8-Paper-Datasets_and_Benchmarks_Track.pdf}
\BIBentrySTDinterwordspacing

\bibitem{debenedetti_agentdojo}
\BIBentryALTinterwordspacing
E.~Debenedetti, J.~Zhang, M.~Balunovi{\'c}, L.~Beurer-Kellner, M.~Fischer, and
  F.~Tram{\`e}r, ``{AgentDojo}: A dynamic environment to evaluate prompt
  injection attacks and defenses for {LLM} agents,'' in \emph{Advances in
  Neural Information Processing Systems}, vol.~37.\hskip 1em plus 0.5em minus
  0.4em\relax Red Hook, NY, USA: Curran Associates, Inc., 2024, pp.
  82\,895--82\,920. [Online]. Available:
  \url{https://proceedings.neurips.cc/paper_files/paper/2024/file/97091a5177d8dc64b1da8bf3e1f6fb54-Paper-Datasets_and_Benchmarks_Track.pdf}
\BIBentrySTDinterwordspacing

\bibitem{stanley_gaap}
\BIBentryALTinterwordspacing
R.~Stanley, A.~Verma, L.~Tsai, K.~Kallas, and S.~Kumar, ``An {AI} agent
  execution environment to safeguard user data,'' 2026. [Online]. Available:
  \url{https://arxiv.org/abs/2604.19657}
\BIBentrySTDinterwordspacing

\bibitem{ghalebikesabi_operationalizing_ci}
\BIBentryALTinterwordspacing
S.~Ghalebikesabi, E.~Bagdasaryan, R.~Yi, I.~Yona, I.~Shumailov, A.~Pappu,
  C.~Shi, L.~Weidinger, R.~Stanforth, L.~Berrada, P.~Kohli, P.-S. Huang, and
  B.~Balle, ``Operationalizing contextual integrity in privacy-conscious
  assistants,'' 2024. [Online]. Available:
  \url{https://arxiv.org/abs/2408.02373}
\BIBentrySTDinterwordspacing

\bibitem{cheng_cibench}
\BIBentryALTinterwordspacing
Z.~Cheng, D.~Wan, M.~Abueg, S.~Ghalebikesabi, R.~Yi, E.~Bagdasarian, B.~Balle,
  S.~Mellem, and S.~O'Banion, ``{CI-Bench}: Benchmarking contextual integrity
  of {AI} assistants on synthetic data,'' 2024. [Online]. Available:
  \url{https://arxiv.org/abs/2409.13903}
\BIBentrySTDinterwordspacing

\bibitem{zharmagambetov_agentdam}
\BIBentryALTinterwordspacing
A.~Zharmagambetov, C.~Guo, I.~Evtimov, M.~Pavlova, R.~Salakhutdinov, and
  K.~Chaudhuri, ``{AgentDAM}: Privacy leakage evaluation for autonomous web
  agents,'' 2025. [Online]. Available: \url{https://arxiv.org/abs/2503.09780}
\BIBentrySTDinterwordspacing

\bibitem{ngong_agentscope}
\BIBentryALTinterwordspacing
I.~C. Ngong, K.~Murugesan, S.~Kadhe, J.~D. Weisz, A.~Dhurandhar, and K.~N.
  Ramamurthy, ``{AgentSCOPE}: Evaluating contextual privacy across agentic
  workflows,'' 2026. [Online]. Available:
  \url{https://arxiv.org/abs/2603.04902}
\BIBentrySTDinterwordspacing

\bibitem{common_criteria_2022}
\BIBentryALTinterwordspacing
{Common Criteria Recognition Arrangement}, ``Common criteria for information
  technology security evaluation, part 1: Introduction and general model,''
  Common Criteria Portal, Tech. Rep. CC:2022 Release 1, 2022. [Online].
  Available:
  \url{https://www.commoncriteriaportal.org/files/ccfiles/CC2022PART1R1.pdf}
\BIBentrySTDinterwordspacing

\bibitem{nist_sp800_53a}
\BIBentryALTinterwordspacing
{Joint Task Force}, ``Assessing security and privacy controls in information
  systems and organizations,'' National Institute of Standards and Technology,
  {NIST} Special Publication 800-53A Revision 5, 2022. [Online]. Available:
  \url{https://csrc.nist.gov/pubs/sp/800/53/a/r5/final}
\BIBentrySTDinterwordspacing

\bibitem{apple_pcc_soc3_report}
{Ernst \& Young LLP} and {Apple Inc.}, ``{SOC 3}: Controls over the {Private
  Cloud Compute Provisioning System},''
  \url{https://help.apple.com/pdf/certifications/en_US/2026-Apple-PCC-SOC-3-Report-05-01-25-to-04-30-26.pdf},
  2026, examination period May 1, 2025--April 30, 2026; issued June 25, 2026;
  accessed September 14, 2026.

\bibitem{gsn_standard}
\BIBentryALTinterwordspacing
{SCSC Assurance Case Working Group}, ``Goal structuring notation community
  standard version 3,'' Safety-Critical Systems Club, Tech. Rep. SCSC-141C,
  2021. [Online]. Available: \url{https://scsc.uk/r1386.pdf}
\BIBentrySTDinterwordspacing

\bibitem{mcgraw_ara}
G.~McGraw, H.~Figueroa, V.~Shepardson, and R.~Bonett, ``An architectural risk
  analysis of machine learning systems: Toward more secure machine learning,''
  \url{https://berryvilleiml.com/docs/ara.pdf}, 2020, {B}erryville Institute of
  Machine Learning, version 1.0.

\bibitem{vicarte_asset_threat}
J.~S. Vicarte, M.~Spoczynski, and M.~Elsaid, ``Threat modeling for {AI}: The
  case for an asset-centric approach,''
  \url{https://arxiv.org/abs/2505.06315v2}, 2025, arXiv:2505.06315v2 [cs.CR].

\bibitem{mitre_atlas}
{MITRE}, ``{ATLAS}: Adversarial threat landscape for artificial-intelligence
  systems,'' \url{https://atlas.mitre.org/}, 2026, living knowledge base,
  accessed September 14, 2026.

\bibitem{databricks_dasf}
{Databricks}, ``Introducing the {Databricks AI Security Framework (DASF)},''
  \url{https://www.databricks.com/blog/introducing-databricks-ai-security-framework-dasf},
  2024, framework introduction, accessed September 14, 2026.

\bibitem{apple_wwdc25_foundation_models}
{Apple Developer}, ``Meet the foundation models framework,''
  \url{https://developer.apple.com/videos/play/wwdc2025/286/}, 2025, {WWDC25}
  video and transcript, accessed July 17, 2026.

\bibitem{apple_wwdc26_foundation_models}
------, ``What's new in the foundation models framework,''
  \url{https://developer.apple.com/videos/play/wwdc2026/241/}, 2026, {WWDC26}
  video and transcript, accessed July 24, 2026.

\bibitem{apple_intelligence_privacy}
{Apple Support}, ``{Apple Intelligence} and privacy on {iPhone},''
  \url{https://support.apple.com/guide/iphone/apple-intelligence-and-privacy-iphe3f499e0e/ios},
  2026, support documentation, accessed May 5, 2026.

\bibitem{apple_foundation_models_policy}
{Apple Developer}, ``Acceptable use requirements for the foundation models
  framework,''
  \url{https://developer.apple.com/apple-intelligence/acceptable-use-requirements-for-the-foundation-models-framework/},
  2026, developer documentation, accessed May 5, 2026.

\bibitem{apple_pcc_security}
{Apple Security Research}, ``Private cloud compute: A new frontier for {AI}
  privacy in the cloud,''
  \url{https://security.apple.com/blog/private-cloud-compute/}, 2024, security
  documentation, accessed May 5, 2026.

\bibitem{apple_pcc_security_guide}
------, ``Private cloud compute security guide,''
  \url{https://security.apple.com/documentation/private-cloud-compute/}, 2026,
  accessed July 17, 2026.

\bibitem{android_aicore_privacy}
{Android Developers Blog}, ``An introduction to privacy and safety for {Gemini
  Nano},''
  \url{https://android-developers.googleblog.com/2024/10/introduction-to-privacy-and-safety-gemini-nano.html},
  2024, developer blog, accessed May 5, 2026.

\bibitem{android_pcc_architecture}
E.~Marchiori, S.~de~Haas, S.~Volnov, R.~Falcon, R.~Pinto, and M.~Zamarato,
  ``Android private compute core architecture,''
  \url{https://arxiv.org/abs/2209.10317}, 2022, arXiv:2209.10317v2.

\bibitem{google_android_aicore_help}
{Google Android Help}, ``About {Android AICore},''
  \url{https://support.google.com/android/answer/17065362}, 2026, accessed May
  7, 2026.

\bibitem{microsoft_recall_click_to_do}
{Microsoft Support}, ``Click to do in recall: Do more with what's on your
  screen,''
  \url{https://support.microsoft.com/en-US/Windows/ai/ai-features/click-to-do-in-recall-do-more-with-what-s-on-your-screen},
  2026, support documentation, accessed July 24, 2026.

\bibitem{microsoft_recall_privacy}
------, ``Privacy and control over your recall experience,''
  \url{https://support.microsoft.com/windows/privacy-and-control-over-your-recall-experience-d404f672-7647-41e5-886c-a3c59680af15},
  2026, support documentation, accessed May 5, 2026.

\bibitem{microsoft_recall_architecture}
{Microsoft Windows Experience Blog}, ``Update on recall security and privacy
  architecture,''
  \url{https://blogs.windows.com/windowsexperience/2024/09/27/update-on-recall-security-and-privacy-architecture/},
  2024, accessed May 7, 2026.

\bibitem{microsoft_recall_snapshots_disk}
{Microsoft Support}, ``Manage your recall snapshots and disk space,''
  \url{https://support.microsoft.com/en-us/windows/ai/ai-features/manage-your-recall-snapshots-and-disk-space},
  2026, accessed July 17, 2026.

\bibitem{google_private_ai_compute}
\BIBentryALTinterwordspacing
J.~Yagnik, ``{Private AI Compute}: Our next step in building private and
  helpful {AI},'' {Google}, 2025, published Nov. 11, 2025. Accessed Sep. 28,
  2026. [Online]. Available:
  \url{https://blog.google/innovation-and-ai/products/google-private-ai-compute/}
\BIBentrySTDinterwordspacing

\bibitem{ncc_private_ai_compute}
\BIBentryALTinterwordspacing
{NCC Group}, ``Public report: {Google Private AI Compute} review,'' 2025,
  report v1.4, Nov. 8, 2025. Accessed Sep. 28, 2026. [Online]. Available:
  \url{https://www.nccgroup.com/research/public-report-google-private-ai-compute-review/}
\BIBentrySTDinterwordspacing

\bibitem{apple_pcc_hardware}
{Apple Security Research}, ``{Private Cloud Compute}: Hardware integrity,''
  \url{https://security.apple.com/documentation/private-cloud-compute/hardwareintegrity},
  2026, security guide, accessed September 14, 2026.

\bibitem{apple_pcc_transparency}
------, ``{Private Cloud Compute}: Release transparency,''
  \url{https://security.apple.com/documentation/private-cloud-compute/releasetransparency},
  2026, security guide, accessed September 14, 2026.

\bibitem{apple_pcc_vre}
------, ``{Private Cloud Compute}: Virtual research environment,''
  \url{https://security.apple.com/documentation/private-cloud-compute/virtualresearchenvironment},
  2026, security guide, accessed September 14, 2026.

\bibitem{apple_pcc_report}
------, ``Appendix: {Apple Intelligence Report},''
  \url{https://security.apple.com/documentation/private-cloud-compute/appendix_appleintelligencereport},
  2026, security guide, accessed September 14, 2026.

\bibitem{apple_pcc_agent}
------, ``{PCC Agent},''
  \url{https://security.apple.com/documentation/private-cloud-compute/pccagent},
  2026, security guide, accessed September 14, 2026.

\bibitem{apple_pcc_transitive}
------, ``{Private Cloud Compute}: Transitive trust,''
  \url{https://security.apple.com/documentation/private-cloud-compute/transitivetrust},
  2026, security guide, accessed September 14, 2026.

\bibitem{apple_pcc_multiturn}
------, ``{Private Cloud Compute}: Multi-turn agent,''
  \url{https://security.apple.com/documentation/private-cloud-compute/multiturnagent},
  2026, security guide, accessed September 14, 2026.

\bibitem{apple_device_management_restrictions}
{Apple Support}, ``Review device management restrictions for apple devices,''
  \url{https://support.apple.com/guide/deployment/review-device-management-restrictions-dep739685973/web},
  2026, accessed May 7, 2026.

\end{thebibliography}

\end{document}